\documentclass[twocolumn,showpacs,prl]{revtex4}

\usepackage{graphicx}
\usepackage{dcolumn}
\usepackage{bm}

\begin{document}

\preprint{APS/123-QED}

\title{Observation of Ultra-narrow Electromagnetically Induced Transparency\\ and Slow Light using Purely Electronic Spins in a Hot Atomic Vapor}

\author{F. Goldfarb$^1$}
\email{Fabienne.Goldfarb@lac.u-psud.fr}
\author{J. Ghosh$^{1,2}$}%
\author{M. David$^1$}%
\author{J. Ruggiero$^1$}%
\author{T. Chaneli\`ere$^1$}%
\author{J.-L. Le Gou\"et$^1$}%
\author{H. Gilles$^3$}%
\author{R. Ghosh$^2$}%
\author{F. Bretenaker$^1$}%

\affiliation{%
$^1$Laboratoire Aim\'e Cotton, CNRS-Universit\'e Paris Sud 11, 91405 Orsay, France\\
$^2$School of Physical Sciences, Jawaharlal Nehru University, New
Delhi - 110 067, India\\
$^3$Centre de Recherche sur les Ions, les Mat\'eriaux et la Photonique, 6 boulevard du Mar\'echal Juin, 14050 Caen, France}

\date{\today}

\begin{abstract}
Electromagnetically induced transparency (EIT) is observed in gaseous $^{4}$He at room temperature. Ultra-narrow (less than $10$ kHz) EIT
windows are obtained for the first time for purely electronic spins in the presence of Doppler broadening. The positive role of collisions
is emphasized through measurements of the power dependence of the EIT resonance. Measurement of slow light opens up possible ways to
applications.
\end{abstract}

\maketitle

Electromagnetically induced transparency (EIT) is a quantum interference effect that permits the propagation of light through an otherwise
opaque medium. A coupling laser creates the interference necessary to allow the transmission of resonant pulses from a probe laser
\cite{boller1991,field1991}. A narrow spectral hole in the absorption profile is accompanied by a strong dispersion of the index of
refraction within the transparency bandwidth, inducing a low group velocity \cite{harris1992,kasapi1995,zibrov1996,hau1999,kash1999}. Other
schemes for low group velocity have been implemented based on again a reduction of the absorption, in coherent population oscillations
\cite{bigelow2003}, and dual absorption lines \cite{camacho2006}, or on a gain resonance, such as stimulated Brillouin scattering
\cite{okawachi2005}, and stimulated Raman scattering \cite{sharping2005}.

Since EIT in three-level $\Lambda$ systems is based on quantum interference effects involving coherence between the two lower states, its
efficiency is strongly dependent on the lifetime of this Raman coherence. This is why the first observations of EIT in hot vapors of Sr
\cite{boller1991} and Pb \cite{field1991}, which involved coherences between two different electronic levels or sublevels, led to very broad
resonances. Very narrow EIT peaks, and consequently, very steep dispersive features could be obtained using cold atom clouds \cite{hau1999}
or low-temperature solids \cite{turukhin2002}. However, for applications requiring methods that can delay a pulse of light in a material
medium in a tunable and controllable fashion \cite{boyd2005,matsko2005,jänes2005,tidström2007}, the quest has been on for a simple,
room-temperature system capable of demonstrating EIT and slow light. Some promising results have been obtained using hot alkali atoms
\cite{zibrov1996,kash1999,budker1999,phillips2001,affolderbach2002,harada2006}. Some sub-kHz EIT features have even been observed using
paraffin-coated cells \cite{klein2006}. However, all these systems use complicated level structures and involve nuclear spins, i.e., the
lower states between which the Raman coherence is built are usually hyperfine sublevels of one electronic sublevel. Collisions are usually
considered as detrimental in these systems, since they destroy the Raman coherences.

In this Letter, we show that metastable $^4$He (He*) is an ideal candidate for ultra-narrow EIT in a $\Lambda$ system \footnote{Notice that
this is very different from the (rather broad) EIT features observed in a ladder system in $^4$He in Ref. \cite{pavone1998}.} involving only
electronic spins in a vapor at room temperature. Indeed, it is already known that it is possible to isolate a perfect $\Lambda$ system in
He* involving only electronic spins and in which the Raman coherence lifetime is limited only by the transit time of the atoms through the
laser beam \cite{gilles2001}. Moreover, we expect collisions to play a favorable role through four different effects involving the
peculiarities of He: (i) velocity-changing collisions enable us to optically pump atoms spanning the entire Doppler profile quickly and
efficiently; (ii) collisions increase the transit time of the atoms through the beam and hence the Raman coherence lifetime
\cite{fitzsimmons1968}; (iii) this is possible because collisions involving He atoms in the zero spin and angular momentum ground state do
not depolarize the colliding He* \cite{phelps1955}; and (iv) Penning ionization among identically polarized He* atoms is almost forbidden
\cite{shlyapnikov1994}.

\begin{figure}
\includegraphics[width=8.5cm]{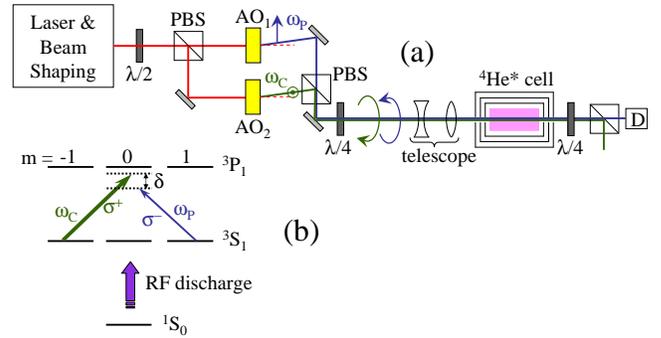}
\caption{\label{fig1} (a) Experimental setup. PBS: polarization beam splitter; D: detector; AO's: acousto-optic modulators. (b) Relevant
level scheme. $\omega_{\mathrm{P}}$ and $\omega_{\mathrm{C}}$: frequencies of the probe and coupling beams, respectively.}
\end{figure}

Fig. \ref{fig1}(a) is a schematic representation of our experimental setup. Light at 1.083 $\mu\mathrm{m}$ resonant with the $^3\mathrm{S}_1
\rightarrow\,^3\mathrm{P}_1$ transition of $^4$He is provided by a diode laser (model SDL-6700). The beam is spatially filtered by passage
through a single-mode fiber. The frequencies and intensities of the coupling and probe beams are adjusted by choosing the amplitudes and the
frequencies of the RF signals driving the two acousto-optic modulators AO$_1$ and AO$_2$. These two beams are recombined and a quarter-wave
plate transforms them into two orthogonal circular polarizations before they enter the helium cell. Different sets of lenses are used at the
entrance of the cell to adjust the Gaussian beam diameter inside the cell between 0.8 and 2.0 cm (at $1/\mathrm{e}^2$ of the maximum
intensity) in order to vary the transit time of the atoms through the beam and consequently the lifetime of the Raman coherences. The
available power for the coupling beam is about 20 mW, which is large enough due to the fact that the saturation intensity in metastable He
is very low (0.16 mW/cm$^2$).

The helium cell is 2.5-cm long and has a diameter of 2.5 cm, and is filled with $^4$He at 1 Torr. He atoms are excited to the metastable
state by an RF discharge at 27 MHz. At the center of the Doppler profile of the optical transition and for a vanishing light intensity, the
cell absorbs about 50\% of the incident intensity, corresponding to a density of $3.5\times10^{10}$ atoms/cm$^{3}$ in the $^3\mathrm{S}_1$
metastable state. These figures of course vary with experimental parameters, such as the RF discharge power. The cell is enclosed in a
three-layer mu-metal magnetic shielding.

In the experiment, when we turn on the $\sigma^+$-polarized coupling beam at frequency $\omega_{\mathrm{C}}$ (see Fig. \ref{fig1}(b)), the
atoms get optically pumped to the $m=+1$ sublevel of the metastable $^3\mathrm{S}_1$ level. Thanks to velocity-changing collisions, we
expect this optical pumping by a narrow linewidth (few MHz) laser to propagate through the entire Doppler profile. When
$\omega_{\mathrm{C}}$ is tuned close to the maximum absorption frequency of the $^3\mathrm{S}_1 \rightarrow\,^3\mathrm{P}_1$ transition, we
measure an optical pumping efficiency of the order of 80\%. We probe the EIT window created by this coupling beam by scanning the frequency
$\omega_{\mathrm{P}}$ of the weak probe beam around $\omega_{\mathrm{C}}$, thus scanning the Raman detuning $\delta/2\pi$ around 0 (see Fig.
\ref{fig1}(a)) between -150 and +150 kHz in 5 ms.
\begin{figure}
\includegraphics[width=8.0cm]{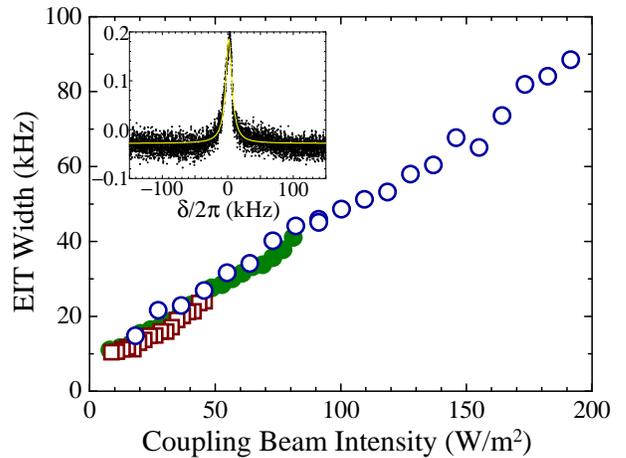}
\caption{\label{fig2} Measured evolution of the EIT window width versus coupling beam intensity for a beam diameter equal to 1 cm (open
circles), 1.5 cm (filled circles) and 2 cm (open squares). The corresponding probe powers are 70 $\mu$W, 100 $\mu$W, and 260 $\mu$W},
respectively. Inset: example of EIT peak: evolution of the logarithm of the measured transmitted probe intensity (in arbitrary units) versus
Raman detuning $\delta$ for a 2.1 mW coupling power and a beam diameter equal to 1.5 cm. The full line is a Lorentzian fit.
\end{figure}
A typical EIT peak is reproduced in the inset of Fig. \ref{fig2}. We choose to calculate the logarithm of the transmitted intensity (as
displayed in Fig. \ref{fig2}) before measuring its full width at half maximum, in order to determine precisely the width of the
susceptibility of the medium. The evolution of this width versus the intensity of the coupling beam is displayed in Fig. \ref{fig2} for
three different beam diameters. To interpret these results, we check with the theory given in Ref. \cite{javan2002} for a Doppler-broadened
medium, in which the role of collisions is completely neglected. The EIT width is thus expected to evolve with the coupling beam Rabi
frequency\footnote{Our Rabi frequency $\Omega_{\mathrm{C}}$ is the real angular frequency of the Rabi oscillations, and is thus double of
the one defined in Refs. \cite{javan2002}, \cite{kuznetsova2002}, and \cite{figueroa2006}.} $\Omega_{\mathrm{C}}$  according to:
\begin{equation}
\Gamma_{\mathrm{EIT}}\simeq\frac{\Omega_{\mathrm{C}}^2}{4\delta_{\mathrm{eff}}}\;, \label{eq1}
\end{equation}
where $\delta_{\mathrm{eff}}$ gives the effective width over which the atoms are pumped into the $m=1$ sublevel of the metastable level for
a fixed value of $\Omega_{\mathrm{C}}$. It is given by $\delta_{\mathrm{eff}}=\Omega_{\mathrm{C}}\sqrt{\Gamma/8\Gamma_{\mathrm{R}}}$ in the
case when $\Omega_{\mathrm{C}} \ll \Omega_{\mathrm{inhom}} = 2\sqrt{2\Gamma_{\mathrm{R}}/\Gamma}W_{\mathrm{D}}$, where $\Gamma$
($\Gamma_{\mathrm{R}}$) is the optical (Raman) coherence decay rate, and $W_{\mathrm{D}}$ is the Doppler half-width at half-maximum. In the
opposite regime, when $\Omega_{\mathrm{C}} \gg \Omega_{\mathrm{inhom}}$, Ref. \cite{javan2002} predicts that
$\delta_{\mathrm{eff}}=W_{\mathrm{D}}$. With our experimental parameters ($\Gamma=1.4\times10^8\,\mathrm{s}^{-1}$ at 1 Torr
\cite{courtade2002}, $\Gamma_{\mathrm{R}}=10^4-10^5\,\mathrm{s}^{-1}$, $W_{\mathrm{D}}/2\pi=0.85$ GHz), we obtain $10^8\,\mathrm{rad/s}\leq
\Omega_{\mathrm{inhom}}\leq 4\times 10^8\,\mathrm{rad/s}$. Since the maximum Rabi frequencies $\Omega_{\mathrm{C}}$ that we reach in our
experiment are smaller than $10^8\,\mathrm{rad/s}$, we are in the first regime where $\Omega_{\mathrm{C}} \ll \Omega_{\mathrm{inhom}}$. We
thus expect $\Gamma_{\mathrm{EIT}}$ to evolve linearly with $\Omega_{\mathrm{C}}$, with a slope depending on $\Gamma_{\mathrm{R}}$
\cite{javan2002}. However, Fig. \ref{fig2} clearly shows that (i) $\Gamma_{\mathrm{EIT}}$ evolves quadratically with $\Omega_{\mathrm{C}}$,
and (ii) the slope of this evolution is the same for different beam sizes, i.e., for different values of $\Gamma_{\mathrm{R}}$. If we use
Eq. (\ref{eq1}) to fit the linear evolution of $\Gamma_{\mathrm{EIT}}$ versus the coupling intensity, we obtain $\delta_{\mathrm{eff}}/2\pi
=\ 0.5\ \mathrm{GHz}$, which is of the same order of magnitude as $W_{\mathrm{D}}/2\pi$, showing that a major part of the Doppler profile
takes part in the EIT process.

This result and the fact that the model of Refs. \cite{javan2002} and \cite{kuznetsova2002} does not fit our measurements, are consistent
with our assumption that velocity-changing collisions are sufficiently efficient to propagate the electronic spin orientation all across the
atomic Doppler profile. This is also consistent with the fact that the mean free path of the He$^*$ atoms is, in a hard sphere model, of the
order of 0.1 mm. If we consider that the atoms cross the beam in a one-dimensional random walk, we can see that, at 300 K, they experience
about $10^4$ collisions during their trip across a 1-cm-diameter beam, leading to a diffusive transit time of the order of 0.5 ms. A more
rigorous calculation using the diffusion constant given in Ref. \cite{fitzsimmons1968} leads to a diffusive transit time of 1 ms through a
1-cm diameter beam at 300 K. The above discussion shows the decisive role played by collisions between metastable and ground state atoms in
our experiments, which is not described by the theories of references \cite{javan2002} and \cite{kuznetsova2002}.

Now, if we suppose that all the atoms across the entire Doppler profile are optically pumped by the coupling beam to the $m = +1$ sublevel
of the metastable level, a calculation of the response of the medium up to first order in probe field leads to the following expression for
the EIT linewidth, as derived in Ref.\,\cite{figueroa2006}:
\begin{equation}
\Gamma_{\mathrm{EIT}}=2\Gamma_{\mathrm{R}}+\frac{\Omega_{\mathrm{C}}^2}{2W_{\mathrm{D}}+\Gamma}\;. \label{eq2}
\end{equation}
The assumption that all the atoms are initially optically pumped thus leads naturally to a linear dependence of the EIT linewidth on the
coupling beam intensity. To obtain Eq.\,\ref{eq2}, one also supposes that the decoherence in the lower states is caused by pure dephasing,
contrary to the assumptions of Refs.\,\cite{javan2002} and \cite{kuznetsova2002}. By fitting the three series of measurements of
Fig.\,\ref{fig2} with straight lines, we obtain linear slopes equal to 380, 410, and 400 $\mathrm{Hz}(\mathrm{W}/\mathrm{m}^2)^{-1}$ for the
2 cm, 1.5 cm, and 1 cm diameter beams, respectively (the measured intensity is averaged over the Gaussian beam profile). This is in very
good agreement with the 416 $\mathrm{Hz}(\mathrm{W}/\mathrm{m}^2)^{-1}$ slope expected from Eq.\,\ref{eq2}. Thus the treatment of
Ref.\,\cite{figueroa2006} quantitatively explains the EIT linewidth results for He*, as it did for rubidium. By extrapolating the
measurements of Fig.\,\ref{fig2} to $\Omega_{\mathrm{C}} = 0$, and using Eq.\,\ref{eq2}, the estimate of $\Gamma_{\mathrm{R}}/2\pi$ comes
out to be 2.8, 3.2 and 4.3 kHz for respectively 2, 1.5 and 1\,cm diameter beam. These values are consistent with the fact that we expect
collisions to increase the transit time of the atoms to the ms range. In the absence of collisions, this transit time would be of the order
of 1 $\mu$s, leading to values of $\Gamma_{\mathrm{R}}$ three orders of magnitude larger than the ones obtained from the measurements of
Fig.\,\ref{fig2}.

\begin{figure}
\includegraphics[width=8.0cm]{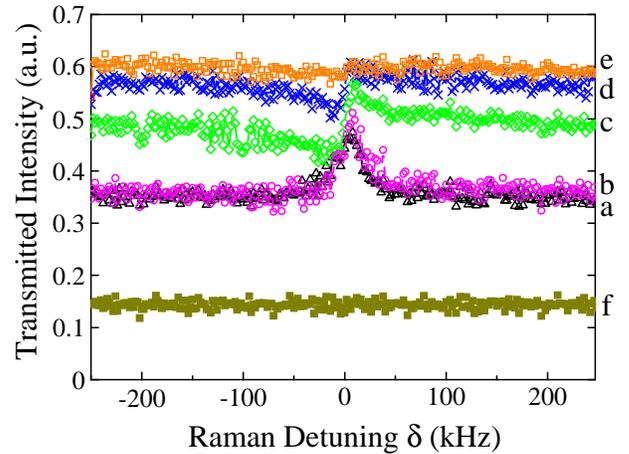}
\caption{\label{fig3} Measured evolution of the transmitted intensity versus Raman detuning $\delta$ for different values of the detuning
$\Delta$ of the coupling beam with respect to the center of the Doppler profile: (a) $\Delta=0$ (triangles), (b) $\Delta=0.4$ GHz (open
circles), (c) $\Delta=1.0$ GHz (diamonds), (d) $\Delta=1.4$ GHz (crosses), (e) $\Delta=2.1$ GHz (open squares), and (f) $\Delta=-2.2$ GHz
(filled squares). All these results have been obtained with a beam diameter of 1.5 cm, a coupling power of 11 mW and a probe power of 140
$\mu$W.}
\end{figure}

To check that the peaks such as the one displayed in the inset of Fig. \ref{fig2} are due to EIT and not any other nonlinear optical
phenomenon, we have recorded such transmitted intensity profiles when the coupling beam frequency is no longer at the center of the Doppler
profile. The corresponding results are displayed in Fig. \ref{fig3} for different values of the coupling beam detuning $\Delta$. We can see
that as soon as $\Delta\neq 0$, the profiles become asymmetric, and may even take a dispersive shape (see, for example, curve (c) obtained
for $\Delta = 1.0$ GHz). This is similar to the Fano profiles obtained in the case of EIT in a homogeneously broadened medium and which have
been shown to be due to interferences between a direct process and stimulated Raman scattering in the overall transition probability
\cite{lounis1992,wong2004}. However, here, these profiles are modified by the fact that they have to be convoluted with the inhomogeneous
Doppler profile. Besides, when we go to negative detunings $\Delta$ (see, for example, the filled squares in Fig. \ref{fig3}), the
transmission is strongly reduced by the absorption due to the neighboring $^3\mathrm{S}_1 \rightarrow\,^3\mathrm{P}_2$ transition, which is
separated from the $^3\mathrm{S}_1 \rightarrow\,^3\mathrm{P}_1$ transition by only 2.29 GHz. The results of Fig. \ref{fig3} show that we are
indeed dealing with EIT, and we have obtained spectral features as narrow as 10 kHz in our system as shown in Fig. \ref{fig2}.

\begin{figure}
\includegraphics[width=8.0cm]{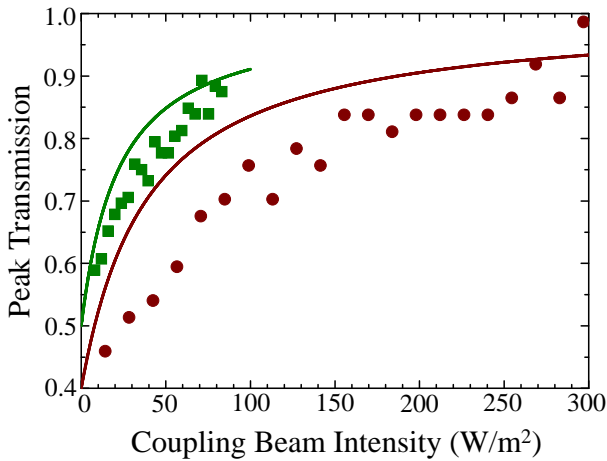}
\caption{\label{fig3bis} Measured evolution of the cell transmission versus coupling beam intensity for two values of the beam diameter: 0.8
cm (circles) and 1.5 cm (squares). The corresponding full lines are obtained using Eq.\,\ref{eq3} with $\Gamma_{\mathrm{R}}/2\pi=$ 5.0 and
3.2 kHz, respectively, and $T_0=$ 0.4 and 0.5 respectively.}
\end{figure}

Another physical parameter that should be studied in this context is the cell transmission $T$. Using the same hypotheses as to derive
Eq.\,\ref{eq2}, its evolution with coupling intensity is predicted to be \cite{figueroa2006}:
\begin{equation}
\ln(T)=\frac{\ln(T_0)}{1+\frac{\Omega_{\mathrm{C}}^2}{2\Gamma_{\mathrm{R}}(2W_{\mathrm{D}}+\Gamma)}}\;. \label{eq3}
\end{equation}
Two examples of such measurements for two beam diameters are reproduced in Fig.\,\ref{fig3bis}. The predictions from Eq.\,\ref{eq3} are the
full lines displayed in Fig.\,\ref{fig3bis} using the values of $\Gamma_{\mathrm{R}}$ extracted from the data of Fig.\,\ref{fig2}. They
reproduce well the shape of the measured evolutions of $T$. Even if the agreement is less good than in Fig. \ref{fig2}, this shows that the
model of Ref.\,\cite{figueroa2006} fairly reproduces the results for EIT in $^4\mathrm{He}^*$.
\begin{figure}
\includegraphics[width=8.0cm]{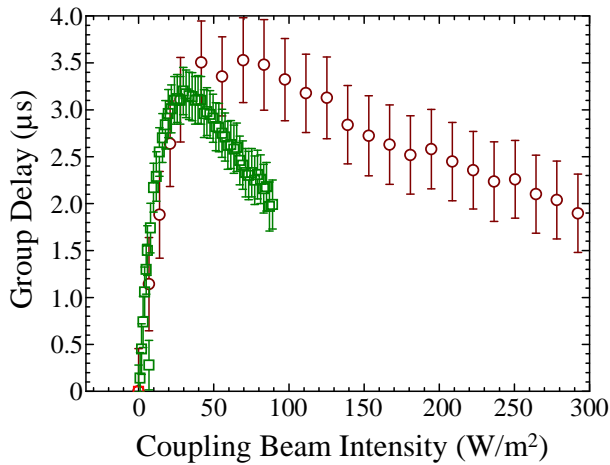}
\caption{\label{fig4} Measured evolution of the group delay through the cell versus coupling beam intensity for two values of the beam
diameter: 0.8 cm (circles) and 1.5 cm (squares). All these results have been obtained with Gaussian probe pulses of duration equal to 70
$\mu$s with a peak power of 35 $\mu$W and with the coupling and probe beam frequencies at the center of the Doppler profile
($\Delta=\delta=0$).}
\end{figure}

\begin{figure}
\includegraphics[width=8.0cm]{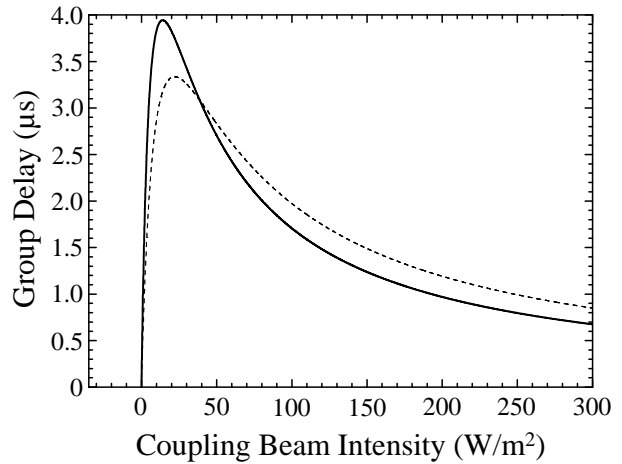}
\caption{\label{fig6} Theoretical evolution of the group delay given by Eq.\,\ref{eq4} with the same sets of parameters as in
Fig.\,\ref{fig3bis}. Full line: $\Gamma_{\mathrm{R}}/2\pi=3.2\ \mathrm{kHz}$ and $T_0=0.5$. Dashed line: $\Gamma_{\mathrm{R}}/2\pi=5.0\
\mathrm{kHz}$ and $T_0=0.4$}
\end{figure}

Narrow spectral features in absorption, such as the one reproduced in the inset of Fig.\,\ref{fig2}, are of course expected to lead to a
strong dispersion and consequently to reduced group velocities. We have checked this by probing our medium with an incident 70-$\mu$s
Gaussian probe pulse tailored by AO$_1$ and measuring the propagation delay through our 2.5-cm long cell in the presence of a coupling beam
at the center of the Doppler profile ($\Delta=0$). The corresponding results are reproduced for two beam diameters in Fig.\,\ref{fig4}. One
can see that group velocities as low as 7000 m/s can be achieved. This corresponds to a maximum delay-bandwidth product of the order of 0.3.
The shape of the evolution of delay versus coupling intensity in Fig. \ref{fig4} can be reproduced using the group velocity derived from the
susceptibility of Ref.\,\cite{figueroa2006}, leading to a group delay at line center ($\Delta=\delta=0$) given by:
\begin{equation}
\tau_{\mathrm{g}}=-\ln(T_0)\frac{(2W_{\mathrm{D}}+\Gamma)\Omega_{\mathrm{C}}^2}{\left[2\Gamma_{\mathrm{R}}(2W_{\mathrm{D}}+\Gamma)
+\Omega_{\mathrm{C}}^2\right]^2}\;. \label{eq4}
\end{equation}
The maximum value of the group delay is reached for $\Omega_{\mathrm{C}}^2=2\Gamma_{\mathrm{R}}(2W_{\mathrm{D}}+\Gamma)$ and is equal to
$-\ln(T_0)/8\Gamma_{\mathrm{R}}$. Using the same parameters as before, Eq.\,\ref{eq4} then leads to the curves of Fig.\,\ref{fig6}. Again,
these curves reproduce the shape of the experimental measurements. In particular, one can see that the maximum achievable delay lies between
3 and 4 $\mu$s, as observed experimentally, and that this maximum delay does not depend strongly on the beam size, as expected from the fact
that $\Gamma_{\mathrm{R}}$ does not depend strongly on the beam size, at least in the range of beam sizes that we have explored.

The preceding results show that the physics of EIT in metastable helium is correctly understood. In particular, the good agreement with a
first order perturbation theory assuming that all the atoms across the Doppler profile are optically pumped by the coupling beam, and the
fact that the Raman coherence lifetimes that we measure are consistent with the transit time of atoms through the beam, provide evidence of
the role of velocity changing collisions in our experiment. However, a fully quantitative understanding is not yet obtained. In particular,
the weakness of the dependence of the Raman coherence lifetime on the beam diameter is not yet quantitatively understood. Such a
quantitative understanding would require a complete treatment of the role of collisions. Moreover, we can expect these collisions to lead to
a dependence of the optical pumping efficiency on the velocity of the atoms and on $\Omega_{\mathrm{C}}$. Similar to what Arimondo
\cite{arimondo1996} has shown in the case of coherent population trapping in the presence of velocity changing collisions, we can also
expect a more rigorous approach to explain the loss of contrast of our EIT peaks with respect to our present first order perturbation
theory, as can be seen from Fig.\,\ref{fig3bis}. We also suspect the relaxation rates of the population and the Raman coherences in the
metastable state to be different. All these features are not taken into account in the present models. Such a complete theory, adapted for
example from Ref. \cite{arimondo1996}, is far beyond the scope of the present paper.

The results of Fig.\,\ref{fig4} may prove interesting for the generation of controllable large bandwidth ($\sim 1$ GHz) delays for radar
applications \cite{tonda2006}. Indeed, the Doppler linewidth of the $^3\mathrm{S}_1 \rightarrow\,^3\mathrm{P}_1$ transition of He* is
compatible with such a large bandwidth, since high power Yb-doped fiber amplifiers are now available at 1.083 $\mu$m in order to broaden the
EIT peaks. These applications usually require delays in the ns to $\mu$s range. The decrease of group delay which would accompany the
broadening of the EIT window in the presence of a multi-Watt coupling beam would lead to useful values of the delays. Moreover, such delays
would be quickly switchable, either via the frequency or power of the coupling beam or via the RF discharge power. Finally, for radar
applications, the 1 $\mu$m wavelength range is particularly favorable since this is precisely the wavelength at which ultra-low intensity
noise semiconductor laser sources are now available \cite{baili2007}: the small size of these lasers together with the small size of the
helium cells we use could lead to potentially useful broadband delay lines.

In conclusion, using metastable $^4$He, we have obtained ultra-narrow (below 10 kHz) EIT window widths for the first time with a system
involving purely electronic spins in a hot atomic vapor. This has been shown to be possible only because of the peculiar properties of
collisions involving metastable helium, namely, the non-depolarizing nature of He* + He collisions, the positive role of velocity-changing
collisions in propagating the atoms' orientation over the Doppler profile, and the fact that Penning ionization is negligible in this case.
We have shown that all these features are not quantitatively described by existing theories of EIT in gases, and that new theoretical
developments should be triggered by the present results. The large group delays observed here open the way to interesting applications in
the domain of broadband delay lines for radars. Finally, the application of the same kind of techniques to the $^3\mathrm{S}_1
\rightarrow\,^3\mathrm{P}_2$ transition should allow us to observe more complicated tripod-like systems \cite{vewinger2003} in a hot vapor.

\acknowledgments The authors wish to thank E. Arimondo, P.-J. Nacher, and M. Pinard for useful discussions. This work is supported by an
Indo-French Networking Project funded by the Department of Science and Technology, Government of India, and the French Ministry of Foreign
Affairs. The work of JG is supported by the Council of Scientific and Industrial Research, India. The stay of RG in France has been
supported by ``C'Nano Ile-de-France'' and ``Triangle de la Physique''. The School of Physical Sciences, Jawaharlal Nehru University, is
supported by the University Grants Commission, India, under a Departmental Research Support scheme.

\bibliography{EITHelium}
\end{document}